\documentclass[aps,pre,twocolumn,amssymb,amsfonts,floatfix,showpacs]{revtex4}

\usepackage{graphicx}
\usepackage{dcolumn}
\usepackage{bm}
\usepackage{mathbbol}
\usepackage[dvips]{color}

\begin{document}

\title{Onset of quantum chaos in one-dimensional bosonic and fermionic systems \\
and its relation to thermalization}

\author{Lea F. Santos}
\email{lsantos2@yu.edu}
\affiliation{Department of Physics, Yeshiva University, New York, New York 10016, USA}
\author{Marcos Rigol}
\email{mrigol@physics.georgetown.edu}
\affiliation{Department of Physics, Georgetown University, Washington, DC 20057, USA}

\begin{abstract}
By means of full exact diagonalization, we study level statistics and the
structure of the eigenvectors of one-dimensional gapless bosonic and fermionic systems 
across the transition from integrability to quantum chaos. These systems are integrable 
in the presence of only nearest-neighbor terms, whereas the addition of next-nearest 
neighbor hopping and interaction may lead to the onset of chaos. We show that the strength 
of the next-nearest neighbor terms required to observe clear signatures of nonintegrability 
is inversely proportional to the system size. Interestingly, the transition to chaos 
is also seen to depend on particle statistics, with bosons responding first to the 
integrability breaking terms. In addition, we discuss the use of delocalization 
measures as main indicators for the crossover from integrability to chaos and the 
consequent viability of quantum thermalization in isolated systems.
\end{abstract}

\maketitle

\section{Introduction}

Random matrix theory (RMT) deals with the statistical properties of ensembles 
of matrices composed of random elements. It was originally designed by Wigner 
in his efforts to understand the statistics of energy levels of nuclei
\cite{Wigner1951} and was further elaborated by several authors, notably  
Mehta \cite{MehtaBook}. RMT received a significant boost with the discovery 
of its connection with classical chaos 
\cite{Casati1980,Berry1981,Zaslavsky1981,BohigasBook}. In particular, it was 
observed that quantum systems whose classical analog are chaotic show the 
same fluctuation properties predicted by RMT. 

The application of RMT was soon extended to the description of other quantum 
many-body systems, such as atoms, molecules, and quantum dots
\cite{HaakeBook,Guhr1998,Alhassid2000,ReichlBook}, and it was not restricted to statistics 
of eigenvalues but accommodated also the analysis of eigenstates  
\cite{Izrailev1990,ZelevinskyRep1996,Kota2001}. Important developments that 
led to the broadening of the theory include the introduction of ensembles of 
random matrices that take into account the predominance of short range 
interactions in real many-body systems \cite{French1970,Brody1981,Flores2001}, the intimate 
connection between quantum transport and spectral properties of mesoscopic 
systems \cite{Thouless1977,Beenakker1997,Guhr1998}, and the relationship between
chaos and quantum thermalization 
\cite{Deutsch1991,Srednicki1994,Flambaum1996,ZelevinskyRep1996,Jacquod1997,Flambaum1997,IzrailevProceed}.

It has been conjectured that the thermalization of finite isolated quantum 
systems is closely related to the onset of chaos and occurs at the level of 
individual states \cite{Deutsch1991,Srednicki1994,rigol08STATc}, which has 
become known as the eigenstate thermalization hypothesis (ETH). Related work 
was done with nuclear shell model calculations and delocalization 
measures \cite{Horoi1995,ZelevinskyRep1996}. More recently, this subject has 
received renewed attention due to its relevance to ultracold gas experiments. 
For example, in a remarkable experiment by a group at Penn State 
\cite{kinoshita06}, it was shown that, after being subject to a strong 
perturbation, a gas of bosons trapped in a (quasi-)one-dimensional geometry
(created by means of a deep two-dimensional optical lattice) did not relax to 
the standard prediction of statistical mechanics. In contrast to those results, 
in another experiment in which a bosonic gas was trapped in a different 
(quasi-)one-dimensional geometry (generated by an atom chip), relaxation 
to a thermal state was inferred to occur in a very short time 
scale \cite{hofferberth07}.

Following those experiments, several theoretical works have explored the question 
of thermalization in nonintegrable isolated quantum systems after a quantum quench 
in one dimension 
\cite{Kollath2007,Manmana2007,rigol09STATa,rigol09STATb,mazets08,cramer08c,roux09,rigol09STATc,roux09a}.
After numerically exploring the nonequilibrium dynamics in finite 
one dimensional (1D) systems, 
thermalization was observed in some regimes \cite{Kollath2007,rigol09STATa,rigol09STATb}
but not in others \cite{Kollath2007,Manmana2007,rigol09STATa,rigol09STATb}, even 
though in all cases integrability was broken. Several factors may play a role in the 
absence of thermalization in finite 1D systems after a quench: (i) the proximity to 
integrable points \cite{rigol09STATa,rigol09STATb}, (ii) the proximity of the energy 
of the initial nonequilibrium state after the quench to the energy of the ground state 
\cite{rigol09STATa,rigol09STATb,roux09a}, (iii) particle statistics and the 
observable considered (in fermionic systems, the momentum distribution function may 
take much longer to relax to equilibrium than other observables \cite{rigol09STATb}); 
and finally (iv) quenching the system across a superfluid/metal to insulator transition 
\cite{Kollath2007,Manmana2007,roux09a}.
Recent numerical studies for bosons and fermions in one dimension have shown that there 
is a direct link between the presence (absence) of thermalization and the validity 
(failure) of the ETH \cite{rigol09STATa,rigol09STATb}.

In the present work, we provide a detailed description of the 
integrable-chaos transition in the one-dimensional bosonic and fermionic 
systems studied in Refs.\ \cite{rigol09STATa} and \cite{rigol09STATb}. These systems are 
clean and have only two-body interactions; the transition to chaos is achieved by 
increasing the strength of next-nearest-neighbor (NNN) terms rather than by adding 
random parameters to the Hamiltonian. Under certain conditions these systems 
may also be mapped onto Heisenberg spin-1/2 chains. Several papers have analyzed spectral 
statistics of disordered \cite{Avishai2002,Santos2004,Kudo2004,Brown2008,Dukesz2009} 
and clean \cite{Hsu1993,Poilblanc1993,Rabson2004,Kudo2005}
1D Heisenberg spin-1/2 systems. Mostly, they were limited to sizes smaller than considered 
here and, in the case of clean systems, focused on properties associated with 
the energy levels, while here 
eigenvectors are also analyzed. Our goal is to establish a direct comparison
between indicators of chaoticity and the results obtained in 
Refs.\ \cite{rigol09STATa} and \cite{rigol09STATb} for thermalization and the validity of ETH.
Our analysis also provides a way to quantify points (i) and (ii) in the 
previous paragraph, which can result in the absence of thermalization in finite 
systems.

Overall, the crossover from integrability to chaos, quantified with spectral 
observables and delocalization measures, mirrors various features of the onset 
of thermalization investigated in Refs.\ \cite{rigol09STATa} and \cite{rigol09STATb},
in particular, the distinct behavior of observables between systems that are 
close and far from integrability, and between eigenstates whose energies are close
and far from the energy of the ground state. We also find that the contrast between 
bosons and fermions pointed out in Ref.\ \cite{rigol09STATb} is translated here 
into the requirement of larger integrability-breaking terms for the onset of 
chaos in fermionic systems. Larger system sizes also facilitate the induction 
of chaos. In addition, we observe that measures of the degree of delocalization 
of eigenstates become smooth functions of energy only in the chaotic regime, 
a behavior that may be used as a signature of chaos.

The paper is organized as follows. Section \ref{Sec:model} describes the model
Hamiltonians studied and their symmetries. Section \ref{Sec:chaos} analyzes the 
integrable-chaos transition based on various quantities. After a brief review 
of the unfolding procedure, Sec.III.A focuses on quantities associated with the 
energy levels, such as level spacing distribution and level number variance. 
Section III.B 
introduces measures of state delocalization, namely information entropy and inverse 
participation ratio (IPR), showing results for the former in the mean field basis. 
Results for the inverse participation ratio and discussions about representations 
are left to the Appendix. Concluding remarks are presented in Sec.\ \ref{Sec:remarks}.

\section{System Model}
\label{Sec:model}

We consider both scenarios: hardcore bosons and spinless fermions
on a periodic one-dimensional lattice in the presence of 
nearest-neighbor (NN) and next-nearest-neighbor (NNN) hopping and 
interaction. The Hamiltonian for bosons $H_B$ and for fermions $H_F$ 
are respectively given by
{\setlength\arraycolsep{0.5pt}
\begin{eqnarray}
&& H_B = \nonumber \\
&& \sum_{i=1}^{L} \left[ -t \left( b_i^{\dagger} b_{i+1} + h.c. \right)
+V \left(n_i^b -\frac{1}{2} \right) \left(n_{i+1}^b
 -\frac{1}{2}\right)
\right.\nonumber \\
&&- \left. t' \left( b_i^{\dagger} b_{i+2} + h.c. \right)
+V' \left(n_{i}^b -\frac{1}{2}\right) \left(n_{i+2}^b -\frac{1}{2}\right)
\right],
\label{bosonHam}
\end{eqnarray}}
and 
{\setlength\arraycolsep{0.5pt}
\begin{eqnarray}
&& H_F = \nonumber \\
&&\sum_{i=1}^{L} \left[ -t \left( f_i^{\dagger} f_{i+1} + h.c. \right)
+V \left(n_i^f -\frac{1}{2} \right) \left(n_{i+1}^f
 -\frac{1}{2}\right)
\right.\nonumber \\
&&- \left. t' \left( f_i^{\dagger} f_{i+2} + h.c. \right)
+V' \left(n_{i}^f -\frac{1}{2}\right) \left(n_{i+2}^f -\frac{1}{2}\right)
\right].
\label{fermionHam}
\end{eqnarray}}

Above, $L$ is the size of the chain, $b_i$ and $b_i^{\dagger}$ 
($f_i$ and $f_i^{\dagger}$) are bosonic (fermionic) annihilation and creation 
operators on site $i$, and $n_i^b= b_i^{\dagger} b_i$ ($n_i^f= f_i^{\dagger} f_i$)
is the boson (fermion) local density operator. Hardcore bosons do not occupy the same 
site, i.e., $b_i^{\dagger 2}=b_i^2$, so the operators commute on different sites 
but can be taken to anti-commute on the same site. The NN (NNN) hopping and interaction 
strengths are respectively $t$ ($t'$) and $V$ ($V'$). Here, we only study repulsive interactions 
($V, V' >0$). We take $\hbar =1$ and $t=V=1$ set the energy scale in the remaining of 
the paper.

The bosonic (fermionic) Hamiltonian conserves the total number of particles $N_b$ ($N_f$) 
and is translational invariant, being therefore composed of independent blocks each 
associated with a total momentum $k$. In the particular case of $k=0$, parity is also 
conserved, and at half-filling, particle-hole symmetry
is present, that is, the bosonic [fermionic] model becomes invariant under 
the transformation $\prod_{i}^{L} (b_i^{\dagger} + b_i)$
[$\prod_{i}^{L} (f_i^{\dagger} + f_i)$], which annihilates particles from filled 
sites and creates them in empty ones. The latter two symmetries will be avoided here 
by selecting $k\neq 0$ and $N_{b,f}=L/3$. For even $L$, we consider $k=1,2, \ldots, L/2-1$ 
and for odd $L$, $k=1,2, \ldots (L-1)/2$. The dimension $D_k$ of each symmetry sector 
studied is given in Table \ref{table:dimensions}.

\begin{table}[h]
\caption{Dimensions of subspaces}
\begin{center}
\begin{tabular}{|c|c|c|c|c|}
\hline \hline 
Bosons &  &  & &  \\
\hline
$L=18$ & $k=1,5,7 $ & $k=2,4,8 $ & $k=3 $ & $k=6 $ \\
\hline
 & 1026 & 1035 & 1028 & 1038 \\
\hline
$L=21$ & $k=7 $ & other $k$'s & &  \\
\hline
  & 5538 & 5537 &  & \\
\hline
$L=24$ & odd $k$'s & $k=2,6,10 $ & $k=4 $ & $k=8 $ \\
\hline
 & 30624 & 30664 & 30666 & 30667 \\
\hline \hline 
Fermions &  &  & &  \\
\hline
$L=18$ & $k=1,5,7 $ & $k=2,4,8 $ & $k=3 $ & $k=6 $ \\
\hline
 & 1035 & 1026 & 1038 & 1028 \\
\hline
$L=21$ & $k=7 $ & other $k$'s & &  \\
\hline
  & 5538 & 5537 &  & \\
\hline
$L=24$ & odd $k$'s & $k=2,6,10 $ & $k=4 $ & $k=8 $ \\
\hline
 & 30624 & 30664 & 30667 & 30666 \\
\hline
\end{tabular}
\end{center}
\label{table:dimensions}
\end{table}

Exact diagonalization is performed to obtain all eigenvalues and eigenvectors
of the systems under investigation. When $t'=V'=0$, models (\ref{bosonHam}) and 
(\ref{fermionHam}) are integrable and may be mapped onto one another via the 
Jordan-Wigner transformation~\cite{Jordan1928}. A correspondence with the Heisenberg 
spin-1/2 chain also holds, in which case the system may be solved with the Bethe 
ansatz~\cite{Bethe1931,Karbach1997}.

\section{Signatures of Quantum Chaos}
\label{Sec:chaos}

The concept of exponential divergence, which is at the heart of classical chaos, 
has no meaning in the quantum domain. Nevertheless, the correspondence 
principle requires that signatures of classical chaos remain in the quantum 
level. Different quantities exist to identify the crossover from the
integrable to the non-integrable regime in quantum systems. We consider both 
spectral observables associated with the eigenvalues and quantities used to 
measure the complexity of the eigenvectors.

\subsection{Spectral observables}

Spectral observables, such as
level spacing distribution and level number variance are investigated below. 
They are intrinsic indicators of the integrable-chaos transition. 
Their analysis are based on the unfolded spectrum of each symmetry sector 
separately.

\subsubsection{Unfolding procedure} 

The procedure of unfolding consists of locally rescaling the energies as follows.
The number of levels with energy less than or equal to a certain value $E$ is 
given by the cumulative spectral function, also known as the staircase function,
$N(E) = \sum_n \Theta (E-E_n)$, where $\Theta$ is the unit step function. 
$N(E)$ has a smooth part $N_{sm}(E)$, which is the cumulative mean level density, 
and a fluctuating part $N_{fl}(E)$, that is, $N(E) =N_{sm}(E) + N_{fl}(E)$. 
Unfolding the spectrum corresponds to mapping the energies 
$\{ E_1, E_2, \ldots, E_D \}$ onto $\{ \epsilon_1, \epsilon_2, \ldots \epsilon_D \}$, 
where $\epsilon_n=N_{sm}(E_n)$, so that the mean level density of the new sequence 
of energies is 1. Different methods are used to separate the smooth part from 
the fluctuating one. Statistics that measure long-range correlations are usually
very sensitive to the adopted unfolding procedure, while short-range correlations 
are less vulnerable \cite{Gomez2002}. Here, we discard 20\% of the energies located 
at the edges of the spectrum, where the fluctuations are large, and obtain
$N_{sm}(E)$ by fitting the staircase function with a polynomial of degree 15.

\subsubsection{Level spacing distribution}

The distribution of spacings $s$ of neighboring energy levels 
\cite{MehtaBook,HaakeBook,Guhr1998,ReichlBook} is the most frequently used 
observable to study short-range fluctuations in the spectrum. Quantum levels of 
integrable systems are not prohibited from crossing and the distribution is 
Poissonian, $P_{ P}(s) = \exp(-s)$. In non-integrable systems, crossings are avoided
and the level spacing distribution is given by the Wigner-Dyson distribution, as
predicted by random matrix theory. 
The form of the Wigner-Dyson distribution depends on the symmetry properties 
of the Hamiltonian. Ensembles of random matrices with time reversal invariance, 
the so-called Gaussian orthogonal ensembles (GOEs), lead to 
$P_{ WD}(s) = (\pi s/2)\exp(-\pi s^2/4)$. 
The same distribution form is achieved for
models  (\ref{bosonHam}) and (\ref{fermionHam}) in the chaotic limit, since they
are also time reversal invariant. However, these systems differ from GOEs in the sense that 
they only have two-body interactions and do not contain random elements. 
Contrary to GOEs and to two-body random ensembles \cite{Brody1981},
the breaking of symmetries here is not caused by randomness, but instead by
the addition of frustrating next-nearest-neighbor couplings.
Notice also that the analysis of level statistics in these systems is 
meaningful only in a particular symmetry sector; if different subspaces are mixed, 
level repulsion may be missed even if the system is chaotic \cite{Kudo2005,Santos2009JMP}.

\begin{figure}[htb]
\includegraphics[width=0.45\textwidth]{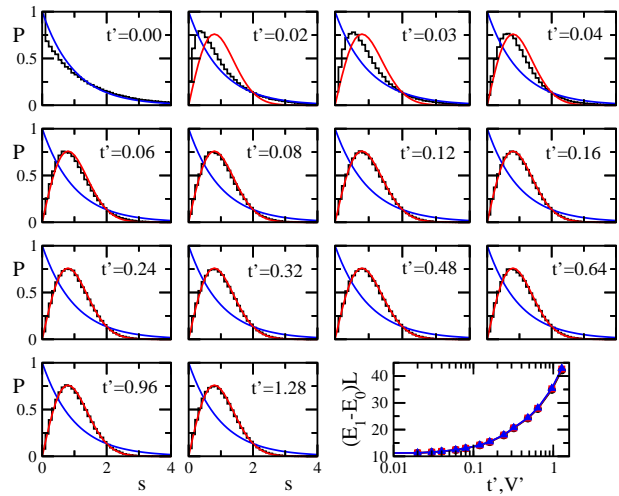}
\vspace{-0.25cm}
\caption{(Color online.) Level spacing distribution for 
hardcore bosons averaged over all $k$'s 
in Table \ref{table:dimensions}, for $L=24$, and $t'=V'$. For comparison purposes, 
we also present the Poisson and Wigner-Dyson distributions. Bottom right panel:
energy difference between first excited state $E_1$ and ground state $E_0$ in the full 
spectrum times $L$, for $L=18$ (circles), $L=21$ (squares), and $L=24$ (triangles).}
\label{fig:bosons}
\end{figure}

\begin{figure}[htb]
\includegraphics[width=0.45\textwidth]{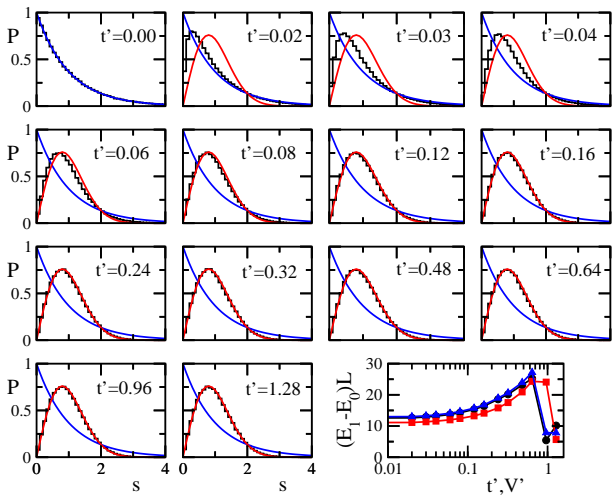}
\vspace{-0.25cm}
\caption{(Color online.) As in Fig.~\ref{fig:bosons} but for spinless fermions.}
\label{fig:fermions}
\end{figure}

In Figs. (\ref{fig:bosons}) and (\ref{fig:fermions}), we show $P(s)$ across 
the transition from integrability to chaos for bosons and fermions, respectively, 
in the case of $L=24$. An average over all $k$'s is performed, but 
we emphasize that the same
behavior is verified also for each $k$-sector separately. 
As $t',V'$ increases and symmetries are broken, level 
repulsion becomes evident, the peak position of the distribution shifts to the
right, and the tail of the distribution changes from exponential to Gaussian. 
Excellent agreement with the Wigner-Dyson distribution is seen already for 
$t'=V'>0.12$. The bottom right panels in Figs.\ (\ref{fig:bosons}) 
and (\ref{fig:fermions}) give the energy difference between first excited 
state and ground state times $L$ as a function of $t',V'$. One can see there 
that the product is size independent emphasizing that the ground state of the 
systems considered here is gapless in the thermodynamic limit, as expected 
from the phase diagrams presented in Ref.\ \cite{zhuravlev97}. Notice that
the particular case of the fermions exhibits an even-odd finite-size effect
that becomes irrelevant in the thermodynamic limit.

\begin{figure}
\includegraphics[width=0.45\textwidth]{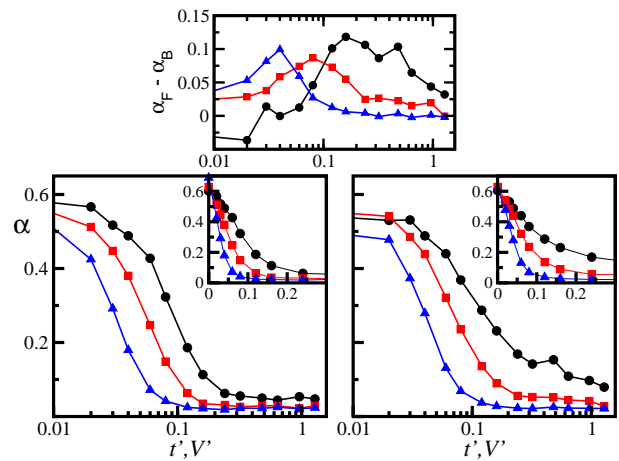}
\vspace{-0.25cm}
\caption{(Color online.) Average $\alpha$ over all $k$'s; $t'=V'$.
Top panel: difference $\alpha$(fermions) - $\alpha$(bosons). Left bottom panel: bosons; 
right bottom panel: fermions. Semi-logarithmic plot in main panels and linear plot in insets.
Circles: $L=18$, squares: $L=21$, triangles: $L=24$.}
\label{fig:eta}
\end{figure}

To better quantify the integrable-chaos transition, we show in Fig.\ \ref{fig:eta}
the level spacing indicator $\alpha$, defined as follows: 
\begin{equation}
\alpha \equiv \frac{\sum_i |P(s_i)-P_{WD}(s_i)|}{\sum_i P_{WD}(s_i)}
\label{eta},
\end{equation}
where the sums runs over the whole spectrum. We should stress that this is a discrete 
rather than integral sum, because $P(s)$ as computed by us is a discrete quantity.
For a chaotic system $\alpha \rightarrow 0$.  The indicator 
$\alpha$ is comparable to the quantity $\eta$ introduced in Ref.\ \cite{Jacquod1997}.

As seen from the bottom panels and insets in Fig.~\ref{fig:eta}, the values of $t',V'$ 
leading to the transition to chaos decrease with the size of the system, suggesting that 
the onset of chaos in the thermodynamic limit might be achieved with an infinitesimally 
small integrability breaking term, although the existence of a saturation value cannot 
be discarded \cite{Rabson2004}. A conclusive statement would require even larger systems 
or a theory for the behavior of systems approaching infinite sizes. Interestingly, 
$\alpha$ decays faster for bosons, which indicates that the crossover to the chaotic 
behavior may depend on particle statistics. This contrasts studies of the ground state 
properties of many-body systems with two-body interactions, where such differences  
were not found \cite{Santos2002}. The top panel in Fig.~\ref{fig:eta} shows the difference 
between the value of $\alpha$ for fermions and bosons. In general, it diminishes with 
increasing the size of the chain, and the point at which the difference attains its 
maximum value moves toward lower values of $t',V'$.

\begin{figure}[htb]
\includegraphics[width=0.45\textwidth]{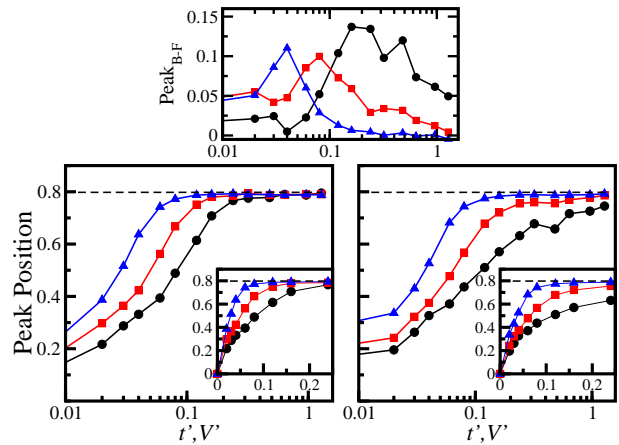}
\vspace{-0.25cm}
\caption{(Color online) Peak position of the level spacing distribution averaged over 
all $k$'s; $t'=V'$. Top panel: difference between the peak position of bosons 
and fermions, peak(bosons) - peak(fermions). Left bottom panel: bosons; right bottom panel: 
fermions. Semi-logarithmic plot in main panels and linear plot in insets. Circles: $L=18$, squares: 
$L=21$, triangles: $L=24$. Dashed line indicates the peak position of $P_{WD}(s)$.}
\label{fig:peak}
\end{figure}

The findings in Fig.~\ref{fig:eta} are reinforced in Fig.~\ref{fig:peak}, where we show the 
approach of the peak position of the level spacing distribution to the peak position of 
$P_{WD}(s)$ as $t',V'$ increases. The transition is faster for larger chains and once again 
depends on particle statistics, with bosons responding first to the breaking of symmetries. 
We should add that equivalent results are obtained by fitting $P(s)$ with the Brody distribution \cite{Brody1981},

\[
P_B(s) = (\beta +1) b s^{\beta} \exp \left( -b s^{\beta +1} \right), \hspace{0.2 cm}
b= \left[\Gamma \left( \frac{\beta + 2}{\beta +1} \right)\right]^{\beta +1},
\]
and analyzing how the increase in $t',V'$ changes $\beta$ from 
0 (in the integrable region) 
to 1 (in the chaotic limit).

The spectral properties of models (\ref{bosonHam}) and (\ref{fermionHam}) discussed 
here are somehow mirrored by their dynamical behavior, which were studied respectively 
in Refs.~\cite{rigol09STATa} and \cite{rigol09STATb}. (Notice that the analysis of the 
integrable-chaos transition performed here uses the same values of $t',V'$ considered 
in those works.) The smooth approach to integrability, as shown in the Figs. 1-4 above, 
is followed by the breakdown of thermalization observed in  Refs.~\cite{rigol09STATa} 
and \cite{rigol09STATb}. However, an important difference between our results here 
and the results in Refs.~\cite{rigol09STATa} and \cite{rigol09STATb}
is that in the latter works it was not clear that the values of $t',V'$ required 
for the system to thermalize would reduce with increasing system size, whereas  
this is the case here for obtaining a Wigner-Dyson distribution of the level spacings.

Another interesting feature found in Ref.\ \cite{rigol09STATb} is that in the context
of quenched dynamics there are differences associated with the particle statistics.
In particular, it was seen that some observables such as the momentum distribution 
function [$n(k)$] in fermionic systems may take longer time to relax to equilibrium than their 
bosonic counterparts. (A related effect in which a quasi-steady regime occurs for $n(k)$ 
before full relaxation has been suggested for higher dimensional fermionic systems 
\cite{moeckel08,moeckel09,Eckstein2009}.) It was also shown in Ref.\ \cite{rigol09STATb} 
that the difference between the eigenstate expectation values of $n(k)$ for eigenstates of 
the fermionic Hamiltonian with close energies suffer 
from particularly large finite size effects when 
compared to other observables such as the density-density structure factor [$N(k)$] and 
when compared to $n(k)$ and $N(k)$ for hardcore bosons. Interestingly, here we find that 
for these finite size systems the measures of chaoticity also exhibit differences between 
hardcore bosons and fermions, where the former ones respond first to the integrability 
breaking terms. Further studies will be required to explore the relation between the latter 
finding and the onset of ETH for different observables and different particle statistics 
in 1D systems.

\subsubsection{Level number variance} 

Other quantities sensitive to spectral fluctuations include measures of 
long-range correlations, such as spectral rigidity and level number variance \cite{Guhr1998}.
Both are closely related and measure the 
deviation of the staircase function from the best fit straight line. Here, we show 
results for the level number variance, $\Sigma^2(l)$, defined as
\begin{equation}
\Sigma^2(l) \equiv \langle (N(l,\epsilon)^2 \rangle - \langle N(l,\epsilon)\rangle^2,
\end{equation}
where $N(l,\epsilon)$ gives the number of states in the interval $[\epsilon,\epsilon+l]$
and $\langle . \rangle $ represents the average over different initial values of $\epsilon$.
For a Poisson distribution, $\Sigma^2(l)=l$, while for GOEs in the limit of large $l$,
$\Sigma^2(l)=2[\ln (2\pi l) + \gamma +1 -\pi^2/8 ]/\pi^2$, where $\gamma$ is the Euler
constant. Level repulsion leads to rather rigid spectra and fluctuations become
much less significant than in the random energy sequences of regular systems.

\begin{figure}[htb]
\includegraphics[width=0.46\textwidth]{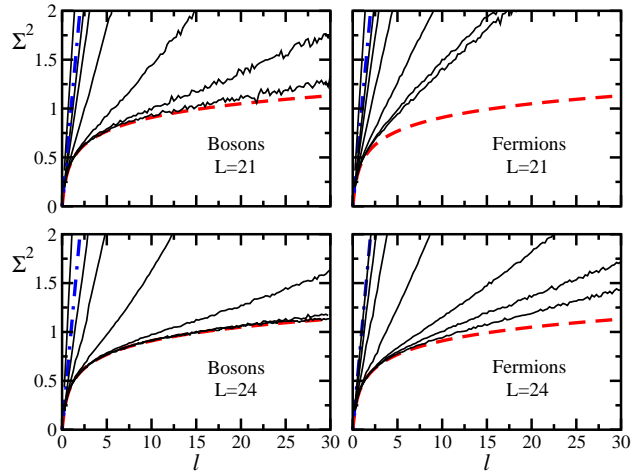}
\vspace{-0.25cm}
\caption{(Color online.) Level number variance averaged over all $k$'s.
In each panel, solid lines from top to bottom: 
$t'=V'=0, 0.02, 0.04, 0.08, 0.16, 0.32, 0.64$. 
Dashed line: GOE, dotted-dashed line: Poisson.}
\label{fig:sigma}
\end{figure}

As expected, the level number variance shown in Fig.~\ref{fig:sigma} approaches 
the GOE curve as the strength of NNN interactions increases. The proximity to the 
GOE result also improves as the system size increases. However, deviations from the 
GOE curve are verified for values of $t',V'$ where the level spacing distribution is 
already very close to a Wigner-Dyson, especially in the case of fermions (right panels).
In fact, the distinct behavior associated with particle statistics becomes yet more 
evident when studying $\Sigma^2(l)$. The level number variance for the bosonic system 
with $L=24$, for example, coincides with the GOE result for a large range of values 
of $l$ already when $t',V'=0.32$ and 0.64, whereas the same is not verified for fermions.
This further supports the view that particle statistics may play an 
important role in the relaxation dynamics and thermalization of finite isolated 
quantum systems. On the other hand, the size dependence of the results is an indicator 
that in the thermodynamic limit the differences between the quantities discussed in 
this paper may become negligible when comparing bosons and fermions, a conjecture that 
deserves further investigation together with its implications for the dynamics and 
thermalization of those systems.

\subsection{Delocalization measures}

Contrary to spectral observables, quantities used to measure the complexity of 
eigenvectors, as delocalization measures \cite{Izrailev1990,ZelevinskyRep1996}, 
are not intrinsic indicators of the 
integrable-chaos transition since they depend on the basis in which the computations 
are performed. The choice of basis is usually physically motivated. The mean-field 
basis is the most appropriate representation to separate global from local properties, 
and therefore capture the transition from regular to chaotic behavior \cite{ZelevinskyRep1996}. 
Here, this basis corresponds to the eigenstates of the integrable Hamiltonian 
($t',V'=0$). Other representations may also provide relevant information, such as the
site basis, which is meaningful in studies of spatial localization, and
the momentum basis, which can be used to study $k$-space localization
(see the Appendix for further discussions).

The degree of complexity of individual eigenvectors may be measured, for example, 
with the information (Shannon) entropy S or the inverse participation ratio (IPR). 
The latter is also sometimes referred to as number of principal components. 
For an eigenstate $\psi_j$ written in the basis vectors $\phi_k$ as 
$\psi_j = \sum_{k=1}^D c^k_j \phi_k$, S and IPR are respectively given by
\begin{equation}
\mbox{S}_j \equiv -\sum_{k=1}^{D}  |c^k_j|^2 \ln |c^k_j|^2,
\label{entropy}
\end{equation}
and
\begin{equation}
\mbox{IPR}_j \equiv \frac{1}{\sum_{k=1}^{D} |c^k_j|^4} .
\label{IPR}
\end{equation}
The above quantities measure the number of basis vectors that contribute to 
each eigenstate, that is, how much delocalized each state is in the chosen basis.

For the GOE, the amplitudes $c_j^k$ are independent random variables and all 
eigenstates are completely delocalized. Complete delocalization does not imply,
however, that S = $\ln D$. For a GOE, the weights $|c^k_j|^2$ fluctuate around 
$1/D$ and the average over the ensemble is 
$\mbox{S}_{\text{GOE}} = \ln(0.48 D) + {\cal O}(1/D)$ 
\cite{Izrailev1990,ZelevinskyRep1996}.

\begin{figure}[htb]
\includegraphics[width=0.47\textwidth]{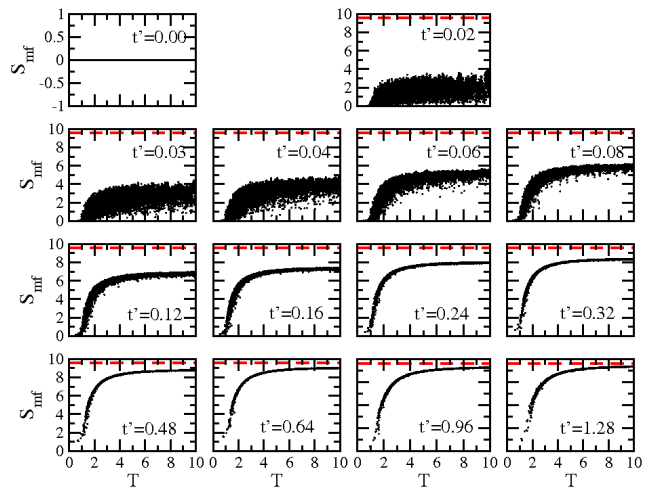}
\vspace{-0.25cm}
\caption{(Color online.) Shannon entropy in the mean field basis 
vs effective temperature for bosons, $L=24$, $k=2$, and $t'=V'$. The dashed line 
gives the GOE averaged value $\mbox{S}_{\text{GOE}} \sim \ln (0.48 D)$.}
\label{fig:Sbosons}
\end{figure}

\begin{figure}[htb]
\includegraphics[width=0.47\textwidth]{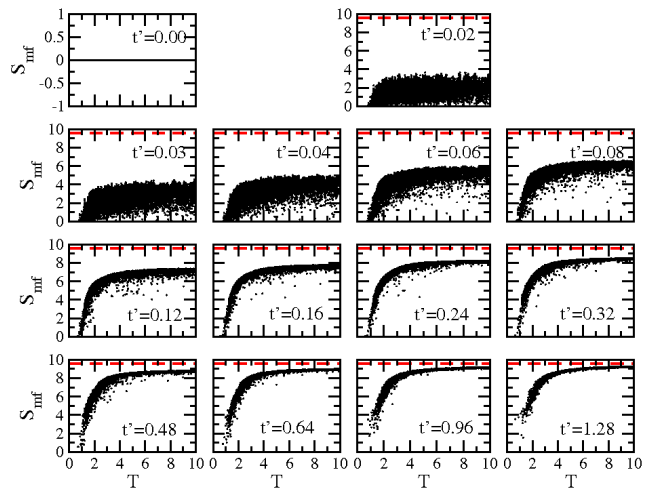}
\vspace{-0.25cm}
\caption{(Color online.) As in Fig.~\ref{fig:Sbosons} for fermions.}
\label{fig:Sfermions}
\end{figure}

Figures \ref{fig:Sbosons} and \ref{fig:Sfermions} show the Shannon entropy in the 
mean field basis S$_{mf}$ vs the effective temperature for bosons and fermions,
respectively. The effective temperature, $T_j$ of an eigenstate $\psi_j$
with energy $E_j$ is defined as
\begin{equation}
E_j = \frac{1}{Z} \mbox{Tr} \left\lbrace \hat{H} e^{-\hat{H}/T_j} \right\rbrace ,
\label{EqTemp}
\end{equation}
where
\begin{equation}
Z=\mbox{Tr} \left\lbrace e^{-\hat{H}/T_j} \right\rbrace .
\end{equation}
Above, $\hat{H}$ is Hamiltonian (\ref{bosonHam}) or (\ref{fermionHam}), $Z$ is
the partition function with the Boltzmann constant $k_B=1$, and the trace is performed over
the full spectrum as in Refs.~\cite{rigol09STATa} and \cite{rigol09STATb}
(see the Appendix for a comparison with effective temperatures obtained by tracing over 
exclusively the sector $k=2$). The figures 
include results only for $T_j \leq 10$; for high $E_j$, the temperatures eventually become 
negative. By plotting the Shannon entropy as a function of the effective temperature, 
we allow for a direct comparison of our results here and the results presented in 
Refs.~\cite{rigol09STATa} and \cite{rigol09STATb}.

As seen in Figs.~\ref{fig:Sbosons} and \ref{fig:Sfermions}, the mixing of basis vectors,
and therefore the complexity of the states, increases with $t',V'$, but it is only for 
$T_j\gtrsim 2$ that the eigenstates of our systems approach the GOE result. Similarly, 
in plots of S$_{mf}$ vs energy (see Figs.~\ref{fig:SmfEBL24} and \ref{fig:SmfEFL24} 
in the Appendix), it is only away from the borders of the spectrum that 
S$_{mf}\rightarrow \mbox{S}_{\text{GOE}}$; in the borders, the states 
are more localized and therefore less ergodic. This feature is typical of systems 
with a finite range of interactions, such as models (\ref{bosonHam}) and (\ref{fermionHam})
and also banded, embedded random matrices, and two-body random ensembles 
\cite{Brody1981,Kaplan2000,Kota2001}. 

The analysis of the structure of the eigenstates hints on what to expect for the dynamics 
of the system. In the context of relaxation dynamics, not only the density of complex
states participating in the dynamics is relevant, but also how similar these states are. 
In close connection, but from a static perspective, it is the onset of chaos that 
guarantees the uniformization of the eigenstates. According to Percival's conjecture
\cite{Percival1973}, the complexity of chaotic wave functions adjacent in energy is 
very similar, they essentially show the same information entropy. A further extension 
of this idea is Berry's conjecture \cite{Berry1977}, which assumes that energy 
eigenfunctions in a time-reversal invariant and ergodic system is a superposition of 
random plane waves. The Eigenstate Thermalization Hypothesis (ETH) \cite{Srednicki1994}
can be related to the validity of Berry's conjecture. ETH states that thermalization 
of an isolated quantum system occurs when each eigenstate already exhibits a thermal 
value for the observables, that is, the eigenstate expectation values do not 
fluctuate between eigenstates close in energy.

In Figs.~\ref{fig:Sbosons} and \ref{fig:Sfermions} 
(see also Figs.~\ref{fig:SmfEBL24}-\ref{fig:SsiteFL24} in the Appendix),
the structure of the eigenstates close in energy reveals fluctuations throughout
the spectrum as we approach integrability, whereas in the chaotic regime, fluctuations 
are mostly restricted to the edges of the spectrum, 
with S$_{mf}$ being a smooth function of 
energy (or temperature) away from the borders. 
A related result was seen in Ref.~\cite{Brown2008}, where a
clear relationship between an entanglement measure and a delocalization
measure for a clean Heisenberg model appeared only in the chaotic limit (two dimensions),
being absent in its integrable counterpart (one dimension).
Fluctuations imply that eigenstates very 
close in energy have different degrees of complexity and localization properties; 
these states may therefore not entirely comply with the ETH. This explains the absence 
of thermalization in the integrable limit. In fact, a similar conclusion was achieved in 
Refs.~\cite{rigol09STATa} and \cite{rigol09STATb} based on plots for the momentum 
distribution function vs energy (Figs. 4 and 7, respectively). There, 
it was observed that fluctuations between expectation values of states close in energy 
increase toward integrability. 

\begin{figure}[htb]
\includegraphics[width=0.46\textwidth]{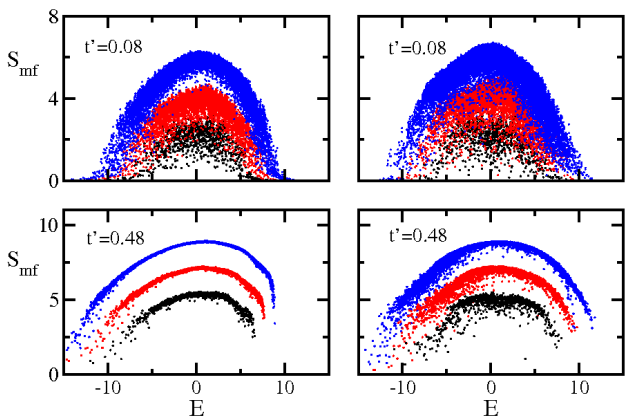}
\vspace{-0.25cm}
\caption{(Color online.) Shannon entropy in the mean field basis
vs energy when the system is close to integrability
(top panels) and in the chaotic limit (bottom panels), 
$k=2$, and $t'=V'$. 
Panels on the left: bosons; panels on the right: fermions.
Curves from bottom to top: $L=18, 21, 24$.}
\label{fig:SvsL}
\end{figure}

Two additional points need to be made here. First, even in the chaotic 
regime one sees that states in the edge of the spectrum remain ``localized'' in the 
mean-field basis. This is accompanied by a failure of ETH in that regime 
\cite{rigol09STATa,rigol09STATb}, and hence thermalization is {\em not expected} to 
occur when the energy of the time-evolving state after the quench is close to 
the ground-state energy (low effective temperatures). This is an important feature 
of isolated quantum systems that will need to be considered with more care when 
dealing with ultracold gases experiments. Second, in comparing bosons and 
fermions, larger fluctuations are verified for the latter, offering a further 
justification for the deviations between statistical mechanics predictions for 
observables after relaxation and the exact time-averaged result of the quantum 
evolution observed in finite fermionic systems \cite{rigol09STATb}. Note, however, 
that fluctuations appear to decrease with system size, as shown in 
Fig.~\ref{fig:SvsL} (specially noticeable in the bottom right panel). 
This may be a simple reflection of better statistics, but may suggest also 
that in the thermodynamic limit some of the differences between fermions and 
bosons may eventually disappear.

\section{Conclusions}
\label{Sec:remarks}

We have presented a detailed analysis of the transition from integrability to 
quantum chaos for gapless one-dimensional systems of interacting spinless fermions and 
hardcore bosons. Here, the onset of chaos was dictated by the enhancement of 
next-nearest-neighbor hopping and interactions.

Our comparisons for different system sizes suggested that in the thermodynamic 
limit an infinitesimal integrability breaking term suffices for the onset of chaos, 
although further studies are necessary for settling this issue. Also, this may not
warrant that thermalization will occur for infinitesimal integrability breaking 
terms since, at least for our finite systems, we could not establish a one to one 
correspondence between the two effects.

We have found differences in behaviors associated with particle statistics.
The transition to chaos in fermionic systems, as measured by level spacing indicator, 
peak position of level spacing distribution, and level number variance, required 
integrability-breaking terms larger than in the bosonic case. With respect to 
delocalization measures, larger fluctuations were also verified for fermions.

We studied wave function complexity using different delocalization measures and
choices of underlying basis. Our results have shown that the similar structure 
of eigenstates close in energy is a primary feature of chaotic systems. This finding 
reinforces the proposal to elevate Berry's conjecture to the status of the best 
definition of quantum chaos \cite{Srednicki1994} and suggests that the onset of 
a smooth dependence of delocalization measures with energy be used as an indicator
of quantum chaos and a condition for quantum thermalization.

Finally, we have shown that even when the systems are chaotic in terms of the 
level spacing distribution and the level number variance, there are still regions 
in the edges of the spectrum in which the states are less delocalized and their
structures less similar. As shown in 
Refs.\ \cite{rigol09STATa} and \cite{rigol09STATb}, those states do not satisfy ETH 
and hence, whenever one performs a quench in a system so that the energy of the time
evolving state is close to the ground state energy [or in other words, when the effective 
temperature of the system as defined by Eq.\ (\ref{EqTemp}) is very low], relaxation
of observables to the thermal distribution prediction is not expected.

The analysis and findings described here are intimately reflected by the studies of
thermalization pursued in Refs.~\cite{rigol08STATc}, \cite{rigol09STATa},
and \cite{rigol09STATb} 
and provide strong support to those works. Moreover, the lattices we considered may 
also be mapped onto other one-dimensional systems, such as spin-1/2 chains,
which indicates the broad range of applicability of our results for gapless systems.

\begin{acknowledgments}
L.F.S. thanks support from the Research Corporation. M.R. was supported by the 
US Office of Naval Research and by Georgetown University.
\end{acknowledgments}

\appendix

\section{Complexity of the wave functions}

We provide here further illustrations for the complexity increase of the wave functions 
with the onset of chaos for models (\ref{bosonHam}) and (\ref{fermionHam}). This is 
based on the computation of Shannon entropy, Eq.~(\ref{entropy}), and the inverse
participation ratio, Eq.~(\ref{IPR}). We compare results in both representations, 
mean-field-  and $k$-space-bases. Overall, the approach to chaos is followed by 
the reduction in fluctuations in the results for S and IPR close in energy, with the 
decrease in fluctuations being slower for fermions than for bosons.

\subsection{Effective temperature}

\begin{figure}[htb]
\includegraphics[width=0.5\textwidth]{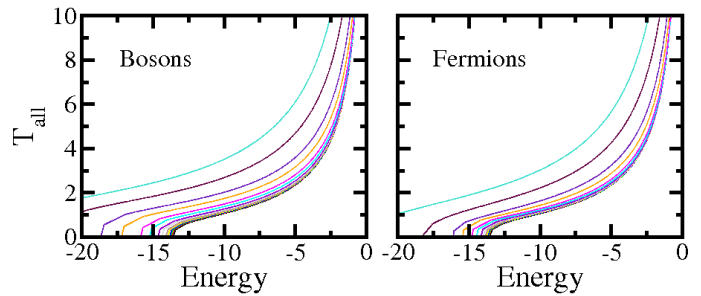}
\vspace{-0.25cm}
\caption{(Color online.) $T_{all}$ vs energy, 
where $T_{all}$ stands for the effective 
temperature computed considering the eigenvalues of all symmetry sectors; 
$L=24$. Curves from bottom to top: 
$t'=V'$ = 0.00, 0.02, 0.03, 0.04, 0.06, 0.08, 0.12, 0.16, 0.24, 0.32, 
0.48, 0.64, 0.96, 1.28.
}
\label{fig:TempvsEner}
\end{figure}

Figures \ref{fig:Sbosons} and \ref{fig:Sfermions} gave the entropy in the mean-field 
basis vs the effective temperature. There, each temperature $T_j$, for an eigenstate 
of energy $E_j$, was obtained by means of Eq.~(\ref{EqTemp}) and performing the trace 
over the full spectrum (let us call it $T_{all}$ here). In
Fig.~\ref{fig:TempvsEner}, we compare $T_{all}$ with the
eigenstate energies. 

\begin{figure}[hb]
\includegraphics[width=0.43\textwidth]{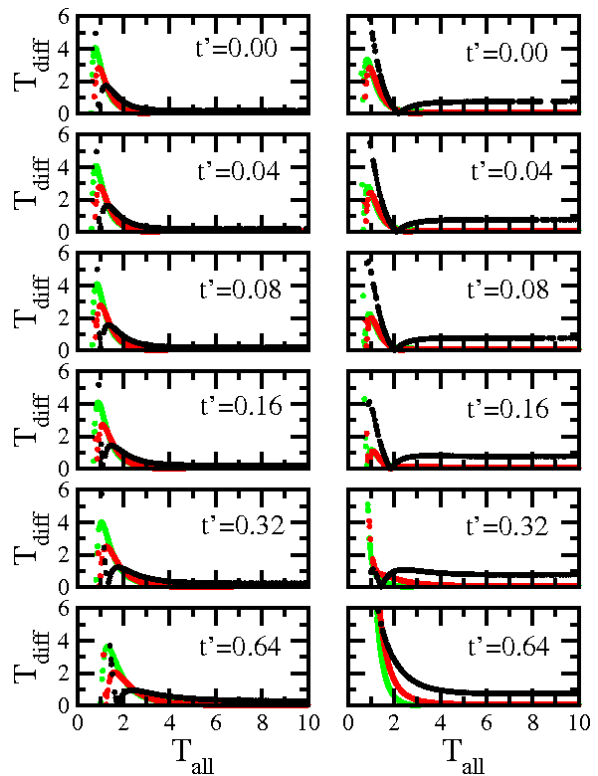}
\vspace{-0.25cm}
\caption{(Color online.)  Temperature difference vs $T_{all}$. 
$T_{diff}=100 |T_{all}-T_{k=2}|/T_{all}$, where 
$T_{k=2}$ is computed considering only the eigenvalues from the $k=2$ sector.
Left panels: bosons; right panels: fermions; $t'=V'$.
Negligible differences are seen between the curves for $L=24$ 
[light gray (green -- online)] and $L=21$ [dark gray (red -- online)]
when $T_{all}>2$, while the curve for $L=18$ (black) saturates at a higher level 
(especially noticeable for fermions).}
\label{fig:Temp}
\end{figure}

One may also wonder what would happen 
if one uses only the spectrum of the $k=2$ sector to perform the trace and hence to 
compute the temperature $T_{k=2}$. Actually, for the system sizes employed here,
the values obtained in this latter way do not differ much from temperatures calculated 
considering the energies of all $k$-sectors \cite{noteTemp}. 
Figure~\ref{fig:Temp} 
shows that the largest disagreements between $T_{all}$ and $T_{k=2}$
occur at low energies; but even then, they 
are usually not higher than 5\%. Exceptions are the first two or three lowest
temperatures, which do not appear in the scale of Fig.~\ref{fig:Temp} due to 
significant discrepancies between $T_{all}$ and $T_{k=2}$.

More generally, for larger system sizes, the temperature is expected to be computed
by means of quantum Monte-Carlo simulations or other better scaling numerical approaches. 
This means that in general all sectors will be considered when computing $T$. Our results 
here show that the differences with considering specific momentum sectors are small and 
decreasing with the system size.

\subsection{Results in the mean-field basis}

\begin{figure}[htb]
\includegraphics[width=0.47\textwidth]{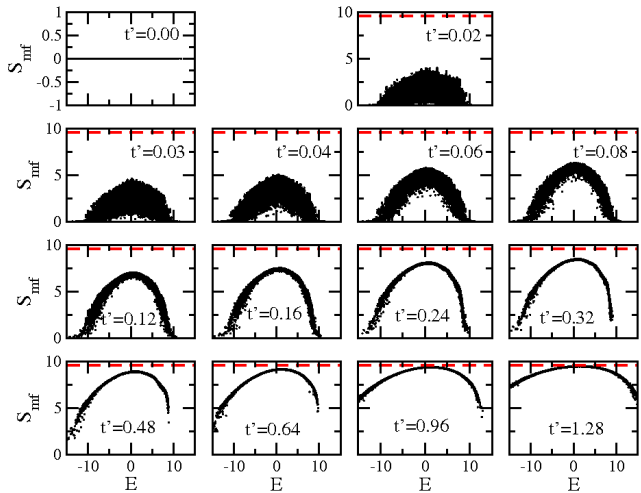}
\vspace{-0.25cm}
\caption{(Color online.)  Shannon entropy in the mean field basis vs
energy for bosons, $L=24$, $k=2$, and $t'=V'$. The dashed line gives the 
GOE averaged value $\mbox{S}_{\text{GOE}} \sim \ln (0.48 D)$.}
\label{fig:SmfEBL24}
\end{figure}

\begin{figure}[htb]
\includegraphics[width=0.47\textwidth]{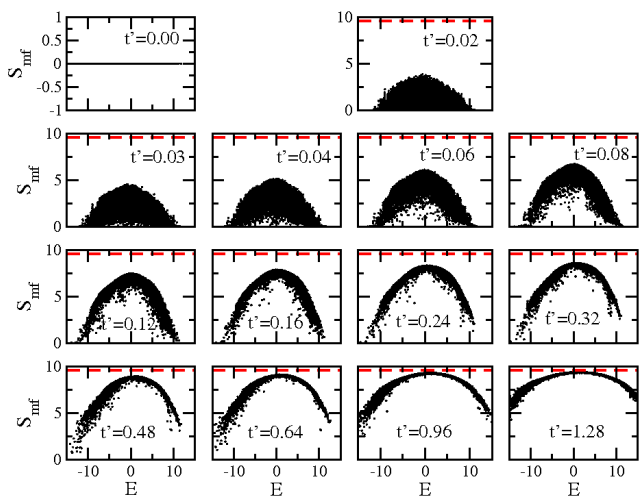}
\vspace{-0.25cm}
\caption{(Color online.) Same as in Fig.~\ref{fig:SmfEBL24}
for fermions.}
\label{fig:SmfEFL24}
\end{figure}

Figures \ref{fig:SmfEBL24} and \ref{fig:SmfEFL24} show the mean-field Shannon entropy vs
energy for all the eigenstates of the $k=2$ sector. Notice that the whole spectrum for 
the $k=2$ sector is presented and not only the energies leading to $T\leq 10$ as in 
Figs.~\ref{fig:Sbosons} and \ref{fig:Sfermions}. The typical behavior of banded matrices 
is observed: larger delocalization appearing away from the edges of the spectrum, 
although not as large as the GOE result 
$\mbox{S}_{\text{GOE}}=\ln(0.48 D) + {\cal O}(1/D)$, 
and lower complexity at the edges~\cite{Brody1981,Kaplan2000,Kota2001}.

\begin{figure}[htb]
\includegraphics[width=0.47\textwidth]{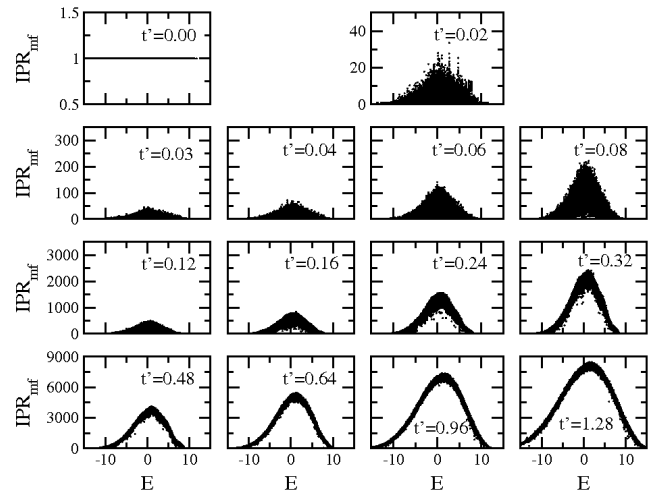}
\vspace{-0.25cm}
\caption{(Color online.) Inverse participation ratio in the mean field basis vs
energy for bosons, $L=24$, $k=2$, and $t'=V'$. The GOE result $\mbox{IPR}_{\text{GOE}} 
\sim D/3$ is beyond the chosen scale.}
\label{fig:IPRmfBL24}
\end{figure}

\begin{figure}[htb]
\includegraphics[width=0.47\textwidth]{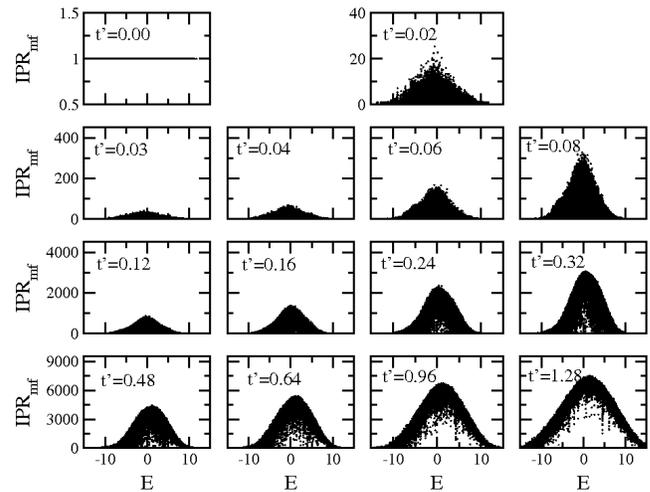}
\vspace{-0.25cm}
\caption{(Color online.) Same as in Fig.~\ref{fig:IPRmfBL24}
for fermions.}
\label{fig:IPRmfFL24}
\end{figure}

A similar behavior is seen in the plots of the inverse participation ratio in the 
mean-field basis vs energy (Figs.~\ref{fig:IPRmfBL24} and \ref{fig:IPRmfFL24}).
The IPR values increase significantly with $t',V'$, but do not reach the GOE result 
$\mbox{IPR} = (D+2)/3$ \cite{Izrailev1990,ZelevinskyRep1996}. IPR gives essentially 
the same information as S, although the first shows larger fluctuations.

\subsection{Results in the $k-$basis}

Identifying the mean-field basis may not always be a simple task. For example,
some 1D models may have more than one integrable point. It may also happen that
one is so far from any integrable point that there is no reason to believe that 
such a point has any relevance for the chosen system. The latter case may be 
particularly applicable to higher-dimensional systems where integrable points
are, in general, the noninteracting limit or other trivial limit. Working on 
the mean-field basis also adds an extra step in the computations since the 
diagonalization of the system is usually not performed in that basis, i.e., 
one needs to perform a change in basis when computing S and IPR in the mean-field 
basis. This extra computation step may become very demanding when dealing with 
large systems. In addition, depending on the studies being performed, it may be 
of interest to analyze the structure of the eigenvectors in another basis. The 
problem of spatial localization, for example, calls for the use of the site-basis.

\begin{figure}[htb]
\includegraphics[width=0.47\textwidth]{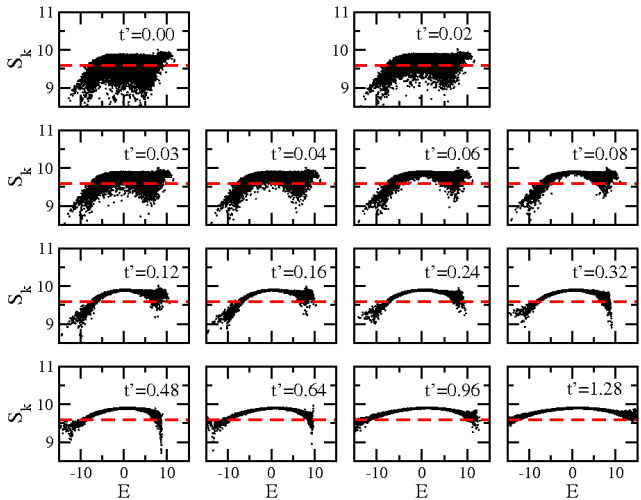}
\vspace{-0.25cm}
\caption{(Color online.) Shannon entropy in the $k$-basis vs
energy for bosons, $L=24$, $k=2$, and $t'=V'$. The dashed line gives the GOE
averaged value $\mbox{S}_{\text{GOE}} \sim \ln (0.48 D)$.}
\label{fig:SsiteBL24}
\end{figure}

\begin{figure}[h!]
\includegraphics[width=0.47\textwidth]{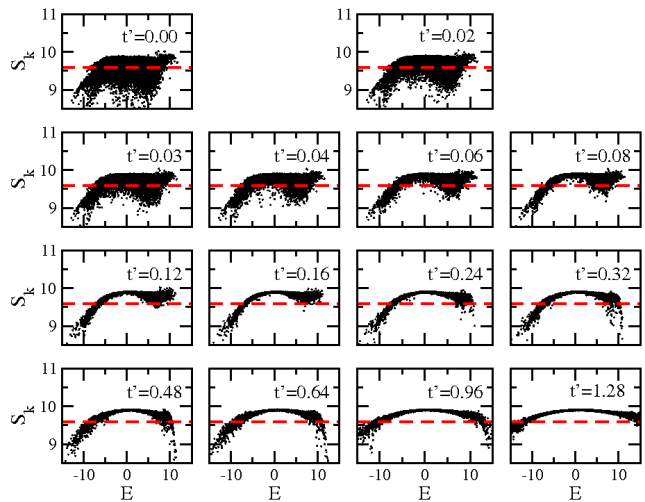}
\vspace{-0.25cm}
\caption{(Color online.) Same as in Fig.~\ref{fig:SsiteBL24}
for fermions.}
\label{fig:SsiteFL24}
\end{figure}

Motivated by the discussion above, we include here the results for S in the 
$k$-basis in which we diagonalize our Hamiltonians. Those are shown in 
Figs.~\ref{fig:SsiteBL24} and \ref{fig:SsiteFL24}. Large entropy values are 
now simply related to high delocalization with respect to the $k$-basis and 
have nothing to do with the onset of chaos. They are found in both integrable 
and non-integrable regimes and may even surpass S$_{\text{GOE}}$. 
Other differences between the mean-field and $k$ bases include:
(i) the localization increase expected for both edges of the spectrum in banded 
matrices is not so evident in the $k$-basis, some high energy states remaining 
as delocalized as the central states; and (ii) the distinct degree of fluctuations 
between bosons and fermions, even though still higher for fermions, 
is not so visible anymore. In spite of these deviations, the $k$-basis may 
still be used as a signature of the integrable-chaos transition. The reason being 
that, just as in the mean field basis, the dependence of S$_k$ 
with energy becomes smoother {\em only} in the chaotic limit. Therefore, since
reduction in fluctuations in S and IPR for states close in energy has been pointed 
as a main cause for the validity of the ETH, the $k$-basis may still be used to 
determine where the onset of thermalization is expected.\\

\vspace{0.5cm}

\begin{thebibliography}{61}
\expandafter\ifx\csname natexlab\endcsname\relax\def\natexlab#1{#1}\fi
\expandafter\ifx\csname bibnamefont\endcsname\relax
  \def\bibnamefont#1{#1}\fi
\expandafter\ifx\csname bibfnamefont\endcsname\relax
  \def\bibfnamefont#1{#1}\fi
\expandafter\ifx\csname citenamefont\endcsname\relax
  \def\citenamefont#1{#1}\fi
\expandafter\ifx\csname url\endcsname\relax
  \def\url#1{\texttt{#1}}\fi
\expandafter\ifx\csname urlprefix\endcsname\relax\def\urlprefix{URL }\fi
\providecommand{\bibinfo}[2]{#2}
\providecommand{\eprint}[2][]{\url{#2}}

\bibitem[{\citenamefont{Wigner}(1951)}]{Wigner1951}
\bibinfo{author}{\bibfnamefont{E.~P.} \bibnamefont{Wigner}},
  \bibinfo{journal}{Ann. Math.} \textbf{\bibinfo{volume}{53}},
  \bibinfo{pages}{36} (\bibinfo{year}{1951}).

\bibitem[{\citenamefont{Mehta}(1991)}]{MehtaBook}
\bibinfo{author}{\bibfnamefont{M.~L.} \bibnamefont{Mehta}},
  \emph{\bibinfo{title}{Random Matrices}} (\bibinfo{publisher}{Academic Press},
  \bibinfo{address}{Boston}, \bibinfo{year}{1991}).

\bibitem[{\citenamefont{Casati et~al.}(1980)\citenamefont{Casati, Valz-Gris,
  and Guarneri}}]{Casati1980}
\bibinfo{author}{\bibfnamefont{G.}~\bibnamefont{Casati}},
  \bibinfo{author}{\bibfnamefont{F.}~\bibnamefont{Valz-Gris}},
  \bibnamefont{and} \bibinfo{author}{\bibfnamefont{I.}~\bibnamefont{Guarneri}},
  \bibinfo{journal}{Lett. Nuov. Cimento Soc. Ital. Fis.} \textbf{\bibinfo{volume}{28}},
  \bibinfo{pages}{279} (\bibinfo{year}{1980}).

\bibitem[{\citenamefont{Berry}(1981)}]{Berry1981}
\bibinfo{author}{\bibfnamefont{M.~V.} \bibnamefont{Berry}},
  \bibinfo{journal}{Ann. Phys. (NY)} \textbf{\bibinfo{volume}{131}},
  \bibinfo{pages}{163} (\bibinfo{year}{1981}).

\bibitem[{\citenamefont{Zaslavsky}(1981)}]{Zaslavsky1981}
\bibinfo{author}{\bibfnamefont{G.~M.} \bibnamefont{Zaslavsky}},
  \bibinfo{journal}{Phys. Rep.} \textbf{\bibinfo{volume}{80}},
  \bibinfo{pages}{157} (\bibinfo{year}{1981}).

\bibitem[{\citenamefont{Bohigas and Giannoni}(1984)}]{BohigasBook}
\bibinfo{author}{\bibfnamefont{O.}~\bibnamefont{Bohigas}} \bibnamefont{and}
  \bibinfo{author}{\bibfnamefont{M.}~\bibnamefont{Giannoni}},
  \emph{\bibinfo{title}{Lecture Notes in Physics vol. 209}}
  (\bibinfo{publisher}{Springer}, \bibinfo{address}{Berlin},
  \bibinfo{year}{1984}).

\bibitem[{\citenamefont{Haake}(1991)}]{HaakeBook}
\bibinfo{author}{\bibfnamefont{F.}~\bibnamefont{Haake}},
  \emph{\bibinfo{title}{Quantum Signatures of Chaos}}
  (\bibinfo{publisher}{Springer-Verlag}, \bibinfo{address}{Berlin},
  \bibinfo{year}{1991}).

\bibitem[{\citenamefont{Guhr et~al.}(1998)\citenamefont{Guhr,
  Mueller-Gr\"oeling, and Weidenm\"uller}}]{Guhr1998}
\bibinfo{author}{\bibfnamefont{T.}~\bibnamefont{Guhr}},
  \bibinfo{author}{\bibfnamefont{A.}~\bibnamefont{Mueller-Gr\"oeling}},
  \bibnamefont{and} \bibinfo{author}{\bibfnamefont{H.~A.}
  \bibnamefont{Weidenm\"uller}}, \bibinfo{journal}{Phys. Rep.}
  \textbf{\bibinfo{volume}{299}}, \bibinfo{pages}{189} (\bibinfo{year}{1998}).

\bibitem[{\citenamefont{Alhassid}(2000)}]{Alhassid2000}
\bibinfo{author}{\bibfnamefont{Y.}~\bibnamefont{Alhassid}},
  \bibinfo{journal}{Rev. Mod. Phys.} \textbf{\bibinfo{volume}{72}},
  \bibinfo{pages}{895} (\bibinfo{year}{2000}).

\bibitem[{\citenamefont{Reichl}(2004)}]{ReichlBook}
\bibinfo{author}{\bibfnamefont{L.~E.} \bibnamefont{Reichl}},
  \emph{\bibinfo{title}{The Transition to Chaos: Conservative Classical Systems
  and Quantum Manifestations}} (\bibinfo{publisher}{Springer},
  \bibinfo{address}{New York}, \bibinfo{year}{2004}).

\bibitem[{\citenamefont{Izrailev}(1990)}]{Izrailev1990}
\bibinfo{author}{\bibfnamefont{F.~M.} \bibnamefont{Izrailev}},
  \bibinfo{journal}{Phys. Rep.} \textbf{\bibinfo{volume}{196}},
  \bibinfo{pages}{299} (\bibinfo{year}{1990}).

\bibitem[{\citenamefont{Zelevinsky et~al.}(1996)\citenamefont{Zelevinsky,
  Brown, Frazier, and Horoi}}]{ZelevinskyRep1996}
\bibinfo{author}{\bibfnamefont{V.}~\bibnamefont{Zelevinsky}},
  \bibinfo{author}{\bibfnamefont{B.~A.} \bibnamefont{Brown}},
  \bibinfo{author}{\bibfnamefont{N.}~\bibnamefont{Frazier}}, \bibnamefont{and}
  \bibinfo{author}{\bibfnamefont{M.}~\bibnamefont{Horoi}},
  \bibinfo{journal}{Phys. Rep.} \textbf{\bibinfo{volume}{276}},
  \bibinfo{pages}{85} (\bibinfo{year}{1996}).

\bibitem[{\citenamefont{Kota}(2001)}]{Kota2001}
\bibinfo{author}{\bibfnamefont{V.~K.~B.} \bibnamefont{Kota}},
  \bibinfo{journal}{Phys. Rep.} \textbf{\bibinfo{volume}{347}},
  \bibinfo{pages}{223} (\bibinfo{year}{2001}).

\bibitem[{\citenamefont{French and Wong}(1970)}]{French1970}
\bibinfo{author}{\bibfnamefont{J.~B.} \bibnamefont{French}} \bibnamefont{and}
  \bibinfo{author}{\bibfnamefont{S.~S.~M.} \bibnamefont{Wong}},
  \bibinfo{journal}{Phys. Lett. B} \textbf{\bibinfo{volume}{33}},
  \bibinfo{pages}{449} (\bibinfo{year}{1970}).

\bibitem[{\citenamefont{Brody et~al.}(1981)\citenamefont{Brody, Flores, French,
  Mello, Pandey, and Wong}}]{Brody1981}
\bibinfo{author}{\bibfnamefont{T.~A.} \bibnamefont{Brody}},
  \bibinfo{author}{\bibfnamefont{J.}~\bibnamefont{Flores}},
  \bibinfo{author}{\bibfnamefont{J.~B.} \bibnamefont{French}},
  \bibinfo{author}{\bibfnamefont{P.~A.} \bibnamefont{Mello}},
  \bibinfo{author}{\bibfnamefont{A.}~\bibnamefont{Pandey}}, \bibnamefont{and}
  \bibinfo{author}{\bibfnamefont{S.~S.~M.} \bibnamefont{Wong}},
  \bibinfo{journal}{Rev. Mod. Phys} \textbf{\bibinfo{volume}{53}},
  \bibinfo{pages}{385} (\bibinfo{year}{1981}).

\bibitem[{\citenamefont{Flores et~al.}(2001)\citenamefont{Flores, Horoi,
  M{\"u}ller, and Seligman}}]{Flores2001}
\bibinfo{author}{\bibfnamefont{J.}~\bibnamefont{Flores}},
  \bibinfo{author}{\bibfnamefont{M.}~\bibnamefont{Horoi}},
  \bibinfo{author}{\bibfnamefont{M.}~\bibnamefont{M{\"u}ller}},
  \bibnamefont{and} \bibinfo{author}{\bibfnamefont{T.~H.}
  \bibnamefont{Seligman}}, \bibinfo{journal}{Phys. Rev. E}
  \textbf{\bibinfo{volume}{63}}, \bibinfo{pages}{026204}
  (\bibinfo{year}{2001}).

\bibitem[{\citenamefont{Thouless}(1977)}]{Thouless1977}
\bibinfo{author}{\bibfnamefont{D.~J.} \bibnamefont{Thouless}},
  \bibinfo{journal}{Phys. Rev. Lett.} \textbf{\bibinfo{volume}{39}},
  \bibinfo{pages}{1167} (\bibinfo{year}{1977}).

\bibitem[{\citenamefont{Beenakker}(1997)}]{Beenakker1997}
\bibinfo{author}{\bibfnamefont{C.~W.~J.} \bibnamefont{Beenakker}},
  \bibinfo{journal}{Rev. Mod. Phys.} \textbf{\bibinfo{volume}{69}},
  \bibinfo{pages}{731} (\bibinfo{year}{1997}).

\bibitem[{\citenamefont{Deutsch}(1991)}]{Deutsch1991}
\bibinfo{author}{\bibfnamefont{J.~M.} \bibnamefont{Deutsch}},
  \bibinfo{journal}{Phys. Rev. A} \textbf{\bibinfo{volume}{43}},
  \bibinfo{pages}{2046} (\bibinfo{year}{1991}).

\bibitem[{\citenamefont{Srednicki}(1994)}]{Srednicki1994}
\bibinfo{author}{\bibfnamefont{M.}~\bibnamefont{Srednicki}},
  \bibinfo{journal}{Phys. Rev. E} \textbf{\bibinfo{volume}{50}},
  \bibinfo{pages}{888} (\bibinfo{year}{1994}).

\bibitem[{\citenamefont{Flambaum et~al.}(1996)\citenamefont{Flambaum, Izrailev,
  and Casati}}]{Flambaum1996}
\bibinfo{author}{\bibfnamefont{V.~V.} \bibnamefont{Flambaum}},
  \bibinfo{author}{\bibfnamefont{F.~M.} \bibnamefont{Izrailev}},
  \bibnamefont{and} \bibinfo{author}{\bibfnamefont{G.}~\bibnamefont{Casati}},
  \bibinfo{journal}{Phys. Rev. E} \textbf{\bibinfo{volume}{54}},
  \bibinfo{pages}{2136} (\bibinfo{year}{1996}).

\bibitem[{\citenamefont{Jacquod and Shepelyansky}(1997)}]{Jacquod1997}
\bibinfo{author}{\bibfnamefont{P.}~\bibnamefont{Jacquod}} \bibnamefont{and}
  \bibinfo{author}{\bibfnamefont{D.~L.} \bibnamefont{Shepelyansky}},
  \bibinfo{journal}{Phys. Rev. Lett.} \textbf{\bibinfo{volume}{79}},
  \bibinfo{pages}{1837} (\bibinfo{year}{1997}).

\bibitem[{\citenamefont{Flambaum and Izrailev}(1997)}]{Flambaum1997}
\bibinfo{author}{\bibfnamefont{V.~V.} \bibnamefont{Flambaum}} \bibnamefont{and}
  \bibinfo{author}{\bibfnamefont{F.~M.} \bibnamefont{Izrailev}},
  \bibinfo{journal}{Phys. Rev. E} \textbf{\bibinfo{volume}{56}},
  \bibinfo{pages}{5144} (\bibinfo{year}{1997}).

\bibitem[{\citenamefont{Izrailev}(2000)}]{IzrailevProceed}
\bibinfo{author}{\bibfnamefont{F.~M.} \bibnamefont{Izrailev}}, in
  \emph{\bibinfo{booktitle}{New Directions in Quantum Chaos}}, 
  Proceedings of the International School of Physics Enrico Fermi No. 143,
  edited by
  \bibinfo{editor}{\bibfnamefont{G.}~\bibnamefont{Casati}},
  \bibinfo{editor}{\bibfnamefont{I.}~\bibnamefont{Guarneri}}, \bibnamefont{and}
  \bibinfo{editor}{\bibfnamefont{U.}~\bibnamefont{Smilansky}}
  (\bibinfo{publisher}{IOS Press}, \bibinfo{address}{Amsterdam},
  \bibinfo{year}{2000}), pp. 371--430.

\bibitem[{\citenamefont{Rigol et~al.}(2008)\citenamefont{Rigol, Dunjko, and
  Olshanii}}]{rigol08STATc}
\bibinfo{author}{\bibfnamefont{M.}~\bibnamefont{Rigol}},
  \bibinfo{author}{\bibfnamefont{V.}~\bibnamefont{Dunjko}}, \bibnamefont{and}
  \bibinfo{author}{\bibfnamefont{M.}~\bibnamefont{Olshanii}},
  \bibinfo{journal}{Nature (London)} \textbf{\bibinfo{volume}{452}},
  \bibinfo{pages}{854} (\bibinfo{year}{2008}).

\bibitem[{\citenamefont{Horoi et~al.}(1995)\citenamefont{Horoi, Zelevinsky, and
  Brown}}]{Horoi1995}
\bibinfo{author}{\bibfnamefont{M.}~\bibnamefont{Horoi}},
  \bibinfo{author}{\bibfnamefont{V.}~\bibnamefont{Zelevinsky}},
  \bibnamefont{and} \bibinfo{author}{\bibfnamefont{B.~A.} \bibnamefont{Brown}},
  \bibinfo{journal}{Phys. Rev. Lett.} \textbf{\bibinfo{volume}{74}},
  \bibinfo{pages}{5194} (\bibinfo{year}{1995}).

\bibitem[{\citenamefont{Kinoshita et~al.}(2006)\citenamefont{Kinoshita, Wenger,
  and Weiss}}]{kinoshita06}
\bibinfo{author}{\bibfnamefont{T.}~\bibnamefont{Kinoshita}},
  \bibinfo{author}{\bibfnamefont{T.}~\bibnamefont{Wenger}}, \bibnamefont{and}
  \bibinfo{author}{\bibfnamefont{D.~S.} \bibnamefont{Weiss}},
  \bibinfo{journal}{Nature (London)} \textbf{\bibinfo{volume}{440}},
  \bibinfo{pages}{900} (\bibinfo{year}{2006}).

\bibitem[{\citenamefont{Hofferberth et~al.}(2007)\citenamefont{Hofferberth,
  Lesanovsky, Fischer, Schumm, and Schmiedmayer}}]{hofferberth07}
\bibinfo{author}{\bibfnamefont{S.}~\bibnamefont{Hofferberth}},
  \bibinfo{author}{\bibfnamefont{I.}~\bibnamefont{Lesanovsky}},
  \bibinfo{author}{\bibfnamefont{B.}~\bibnamefont{Fischer}},
  \bibinfo{author}{\bibfnamefont{T.}~\bibnamefont{Schumm}}, \bibnamefont{and}
  \bibinfo{author}{\bibfnamefont{J.}~\bibnamefont{Schmiedmayer}},
  \bibinfo{journal}{Nature (London)} \textbf{\bibinfo{volume}{449}},
  \bibinfo{pages}{324} (\bibinfo{year}{2007}).

\bibitem[{\citenamefont{Kollath et~al.}(2007)\citenamefont{Kollath,
  L{\"a}uchli, and Altman}}]{Kollath2007}
\bibinfo{author}{\bibfnamefont{C.}~\bibnamefont{Kollath}},
  \bibinfo{author}{\bibfnamefont{A.~M.} \bibnamefont{L{\"a}uchli}},
  \bibnamefont{and} \bibinfo{author}{\bibfnamefont{E.}~\bibnamefont{Altman}},
  \bibinfo{journal}{Phys. Rev. Lett.} \textbf{\bibinfo{volume}{98}},
  \bibinfo{pages}{180601} (\bibinfo{year}{2007}).

\bibitem[{\citenamefont{Manmana et~al.}(2007)\citenamefont{Manmana, Wessel,
  Noack, and Muramatsu}}]{Manmana2007}
\bibinfo{author}{\bibfnamefont{S.~R.} \bibnamefont{Manmana}},
  \bibinfo{author}{\bibfnamefont{S.}~\bibnamefont{Wessel}},
  \bibinfo{author}{\bibfnamefont{R.~M.} \bibnamefont{Noack}}, \bibnamefont{and}
  \bibinfo{author}{\bibfnamefont{A.}~\bibnamefont{Muramatsu}},
  \bibinfo{journal}{Phys. Rev. Lett.} \textbf{\bibinfo{volume}{98}},
  \bibinfo{pages}{210405} (\bibinfo{year}{2007}).

\bibitem[{\citenamefont{Rigol}(2009)}]{rigol09STATa}
\bibinfo{author}{\bibfnamefont{M.}~\bibnamefont{Rigol}},
  \bibinfo{journal}{Phys. Rev. Lett.} \textbf{\bibinfo{volume}{103}},
  \bibinfo{pages}{100403} (\bibinfo{year}{2009}).

\bibitem[{\citenamefont{Rigol}({\natexlab{a}})}]{rigol09STATb}
\bibinfo{author}{\bibfnamefont{M.}~\bibnamefont{Rigol}},
  \bibinfo{journal}{Phys. Rev. A} \textbf{\bibinfo{volume}{80}},
  \bibinfo{pages}{053607} (\bibinfo{year}{2009}).

\bibitem[{\citenamefont{Mazets et~al.}(2008)\citenamefont{Mazets, Schumm, and
  Schmiedmayer}}]{mazets08}
\bibinfo{author}{\bibfnamefont{I.~E.} \bibnamefont{Mazets}},
  \bibinfo{author}{\bibfnamefont{T.}~\bibnamefont{Schumm}}, \bibnamefont{and}
  \bibinfo{author}{\bibfnamefont{J.}~\bibnamefont{Schmiedmayer}},
  \bibinfo{journal}{Phys. Rev. Lett.} \textbf{\bibinfo{volume}{100}},
  \bibinfo{pages}{210403} (\bibinfo{year}{2008}).

\bibitem[{\citenamefont{Flesch et~al.}(2008)\citenamefont{Flesch, Cramer,
  McCulloch, Schollw\"ock, and Eisert}}]{cramer08c}
\bibinfo{author}{\bibfnamefont{A.}~\bibnamefont{Flesch}},
  \bibinfo{author}{\bibfnamefont{M.}~\bibnamefont{Cramer}},
  \bibinfo{author}{\bibfnamefont{I.~P.} \bibnamefont{McCulloch}},
  \bibinfo{author}{\bibfnamefont{U.}~\bibnamefont{Schollw\"ock}},
  \bibnamefont{and} \bibinfo{author}{\bibfnamefont{J.}~\bibnamefont{Eisert}},
  \bibinfo{journal}{Phys. Rev. A} \textbf{\bibinfo{volume}{78}},
  \bibinfo{pages}{033608} (\bibinfo{year}{2008}).

\bibitem[{\citenamefont{Roux}(2009)}]{roux09}
\bibinfo{author}{\bibfnamefont{G.}~\bibnamefont{Roux}}, \bibinfo{journal}{Phys.
  Rev. A} \textbf{\bibinfo{volume}{79}}, \bibinfo{pages}{021608(R)}
  (\bibinfo{year}{2009}).

\bibitem[{\citenamefont{Rigol}({\natexlab{b}})}]{rigol09STATc}
\bibinfo{author}{\bibfnamefont{M.}~\bibnamefont{Rigol}},
  \bibinfo{howpublished}{arXiv:0909.4556}.

\bibitem[{\citenamefont{Roux}()}]{roux09a}
\bibinfo{author}{\bibfnamefont{G.}~\bibnamefont{Roux}},
  \bibinfo{note}{arXiv:0909.4620}.

\bibitem[{\citenamefont{Avishai et~al.}(2002)\citenamefont{Avishai, Richert,
  and Berkovits}}]{Avishai2002}
\bibinfo{author}{\bibfnamefont{Y.}~\bibnamefont{Avishai}},
  \bibinfo{author}{\bibfnamefont{J.}~\bibnamefont{Richert}}, \bibnamefont{and}
  \bibinfo{author}{\bibfnamefont{R.}~\bibnamefont{Berkovitz}},
  \bibinfo{journal}{Phys. Rev. B} \textbf{\bibinfo{volume}{66}},
  \bibinfo{pages}{052416} (\bibinfo{year}{2002}).

\bibitem[{\citenamefont{Santos}(2004)}]{Santos2004}
\bibinfo{author}{\bibfnamefont{L.~F.} \bibnamefont{Santos}},
  \bibinfo{journal}{J. Phys. A} \textbf{\bibinfo{volume}{37}},
  \bibinfo{pages}{4723} (\bibinfo{year}{2004}).

\bibitem[{\citenamefont{Kudo and Deguchi}(2004)}]{Kudo2004}
\bibinfo{author}{\bibfnamefont{K.}~\bibnamefont{Kudo}} \bibnamefont{and}
  \bibinfo{author}{\bibfnamefont{T.}~\bibnamefont{Deguchi}},
  \bibinfo{journal}{Phys. Rev. B} \textbf{\bibinfo{volume}{69}},
  \bibinfo{pages}{132404} (\bibinfo{year}{2004}).

\bibitem[{\citenamefont{Brown et~al.}(2008)\citenamefont{Brown, Santos,
  Starling, and Viola}}]{Brown2008}
\bibinfo{author}{\bibfnamefont{W.~G.} \bibnamefont{Brown}},
  \bibinfo{author}{\bibfnamefont{L.~F.} \bibnamefont{Santos}},
  \bibinfo{author}{\bibfnamefont{D.~J.}~\bibnamefont{Starling}}, \bibnamefont{and}
  \bibinfo{author}{\bibfnamefont{L.}~\bibnamefont{Viola}},
  \bibinfo{journal}{Phys. Rev. E} \textbf{\bibinfo{volume}{77}},
  \bibinfo{pages}{021106} (\bibinfo{year}{2008}).

\bibitem[{\citenamefont{Dukesz et~al.}(2009)\citenamefont{Dukesz, Zilbergerts,
  and Santos}}]{Dukesz2009}
\bibinfo{author}{\bibfnamefont{F.}~\bibnamefont{Dukesz}},
  \bibinfo{author}{\bibfnamefont{M.}~\bibnamefont{Zilbergerts}},
  \bibnamefont{and} \bibinfo{author}{\bibfnamefont{L.~F.}
  \bibnamefont{Santos}}, \bibinfo{journal}{New J. Phys.}
  \textbf{\bibinfo{volume}{11}}, \bibinfo{pages}{043026}
  (\bibinfo{year}{2009}).

\bibitem[{\citenamefont{Hsu and Angles d'Auriac}(1993)}]{Hsu1993}
\bibinfo{author}{\bibfnamefont{T.~C.} \bibnamefont{Hsu}} \bibnamefont{and}
  \bibinfo{author}{\bibfnamefont{J.~C.} \bibnamefont{Angles d'Auriac}},
  \bibinfo{journal}{Phys. Rev. B} \textbf{\bibinfo{volume}{47}},
  \bibinfo{pages}{14291} (\bibinfo{year}{1993}).

\bibitem[{\citenamefont{Poilblanc et~al.}(1993)\citenamefont{Poilblanc, Ziman,
  Bellissard, Mila, and Montambaux}}]{Poilblanc1993}
\bibinfo{author}{\bibfnamefont{D.}~\bibnamefont{Poilblanc}},
  \bibinfo{author}{\bibfnamefont{T.}~\bibnamefont{Ziman}},
  \bibinfo{author}{\bibfnamefont{J.}~\bibnamefont{Bellissard}},
  \bibinfo{author}{\bibfnamefont{F.}~\bibnamefont{Mila}}, \bibnamefont{and}
  \bibinfo{author}{\bibfnamefont{G.}~\bibnamefont{Montambaux}},
  \bibinfo{journal}{EPL} \textbf{\bibinfo{volume}{22}},
  \bibinfo{pages}{537} (\bibinfo{year}{1993}).

\bibitem[{\citenamefont{Rabson et~al.}(2004)\citenamefont{Rabson, Narozhny, and
  Millis}}]{Rabson2004}
\bibinfo{author}{\bibfnamefont{D.~A.} \bibnamefont{Rabson}},
  \bibinfo{author}{\bibfnamefont{B.~N.} \bibnamefont{Narozhny}},
  \bibnamefont{and} \bibinfo{author}{\bibfnamefont{A.~J.}
  \bibnamefont{Millis}}, \bibinfo{journal}{Phys. Rev. B}
  \textbf{\bibinfo{volume}{69}}, \bibinfo{pages}{054403}
  (\bibinfo{year}{2004}).

\bibitem[{\citenamefont{Kudo and Deguchi}(2005)}]{Kudo2005}
\bibinfo{author}{\bibfnamefont{K.}~\bibnamefont{Kudo}} \bibnamefont{and}
  \bibinfo{author}{\bibfnamefont{T.}~\bibnamefont{Deguchi}},
  \bibinfo{journal}{J. Phys. Soc. Jpn.} \textbf{\bibinfo{volume}{74}},
  \bibinfo{pages}{1992} (\bibinfo{year}{2005}).

\bibitem[{\citenamefont{Jordan and Wigner}(1928)}]{Jordan1928}
\bibinfo{author}{\bibfnamefont{P.}~\bibnamefont{Jordan}} \bibnamefont{and}
  \bibinfo{author}{\bibfnamefont{E.}~\bibnamefont{Wigner}},
  \bibinfo{journal}{Z. Phys.} \textbf{\bibinfo{volume}{47}},
  \bibinfo{pages}{631} (\bibinfo{year}{1928}).

\bibitem[{\citenamefont{Bethe}(1931)}]{Bethe1931}
\bibinfo{author}{\bibfnamefont{H.~A.} \bibnamefont{Bethe}},
  \bibinfo{journal}{Z. Phys.} \textbf{\bibinfo{volume}{71}},
  \bibinfo{pages}{205} (\bibinfo{year}{1931}).

\bibitem[{\citenamefont{Karbach and M\"uller}(1997)}]{Karbach1997}
\bibinfo{author}{\bibfnamefont{M.}~\bibnamefont{Karbach}} \bibnamefont{and}
  \bibinfo{author}{\bibfnamefont{G.}~\bibnamefont{M\"uller}},
  \bibinfo{journal}{Comput. Phys.} \textbf{\bibinfo{volume}{11}},
  \bibinfo{pages}{36} (\bibinfo{year}{1997}).

\bibitem[{\citenamefont{G\'omez et~al.}(2002)\citenamefont{G\'omez, Molina,
  Relano, and Retamosa}}]{Gomez2002}
\bibinfo{author}{\bibfnamefont{J.~M.~G.} \bibnamefont{G\'omez}},
  \bibinfo{author}{\bibfnamefont{R.~A.} \bibnamefont{Molina}},
  \bibinfo{author}{\bibfnamefont{A.}~\bibnamefont{Relano}}, \bibnamefont{and}
  \bibinfo{author}{\bibfnamefont{J.}~\bibnamefont{Retamosa}},
  \bibinfo{journal}{Phys. Rev. E} \textbf{\bibinfo{volume}{66}},
  \bibinfo{pages}{036209} (\bibinfo{year}{2002}).

\bibitem[{\citenamefont{Santos}(2009)}]{Santos2009JMP}
\bibinfo{author}{\bibfnamefont{L.~F.} \bibnamefont{Santos}},
  \bibinfo{journal}{J. Math. Phys} \textbf{\bibinfo{volume}{50}},
  \bibinfo{pages}{095211} (\bibinfo{year}{2009}).

\bibitem[{\citenamefont{Zhuravlev et~al.}(1997)\citenamefont{Zhuravlev,
  Katsnelson, and Trefilov}}]{zhuravlev97}
\bibinfo{author}{\bibfnamefont{A.~K.} \bibnamefont{Zhuravlev}},
  \bibinfo{author}{\bibfnamefont{M.~I.} \bibnamefont{Katsnelson}},
  \bibnamefont{and} \bibinfo{author}{\bibfnamefont{A.~V.}
  \bibnamefont{Trefilov}}, \bibinfo{journal}{Phys. Rev. B}
  \textbf{\bibinfo{volume}{56}}, \bibinfo{pages}{12939} (\bibinfo{year}{1997}).

\bibitem[{\citenamefont{Santos et~al.}(2002)\citenamefont{Santos, Kusnezov, and
  Jacquod}}]{Santos2002}
\bibinfo{author}{\bibfnamefont{L.~F.} \bibnamefont{Santos}},
  \bibinfo{author}{\bibfnamefont{D.}~\bibnamefont{Kusnezov}}, \bibnamefont{and}
  \bibinfo{author}{\bibfnamefont{P.}~\bibnamefont{Jacquod}},
  \bibinfo{journal}{Phys. Lett. B.} \textbf{\bibinfo{volume}{537}},
  \bibinfo{pages}{62} (\bibinfo{year}{2002}).

\bibitem[{\citenamefont{Moeckel and Kehrein}(2008)}]{moeckel08}
\bibinfo{author}{\bibfnamefont{M.}~\bibnamefont{Moeckel}} \bibnamefont{and}
  \bibinfo{author}{\bibfnamefont{S.}~\bibnamefont{Kehrein}},
  \bibinfo{journal}{Phys. Rev. Lett.} \textbf{\bibinfo{volume}{100}},
  \bibinfo{pages}{175702} (\bibinfo{year}{2008}).

\bibitem[{\citenamefont{Moeckel and Kehrein}(2009)}]{moeckel09}
\bibinfo{author}{\bibfnamefont{M.}~\bibnamefont{Moeckel}} \bibnamefont{and}
  \bibinfo{author}{\bibfnamefont{S.}~\bibnamefont{Kehrein}},
  \bibinfo{journal}{Ann. Phys.} \textbf{\bibinfo{volume}{324}},
  \bibinfo{pages}{2146} (\bibinfo{year}{2009}).

\bibitem[{\citenamefont{Eckstein et~al.}(2009)\citenamefont{Eckstein, Kollar,
  and Werner}}]{Eckstein2009}
\bibinfo{author}{\bibfnamefont{M.}~\bibnamefont{Eckstein}},
  \bibinfo{author}{\bibfnamefont{M.}~\bibnamefont{Kollar}}, \bibnamefont{and}
  \bibinfo{author}{\bibfnamefont{P.}~\bibnamefont{Werner}},
  \bibinfo{journal}{Phys. Rev. Lett.} \textbf{\bibinfo{volume}{103}},
  \bibinfo{pages}{056403} (\bibinfo{year}{2009}).

\bibitem[{\citenamefont{Kaplan and Papenbrock}(2000)}]{Kaplan2000}
\bibinfo{author}{\bibfnamefont{L.}~\bibnamefont{Kaplan}} \bibnamefont{and}
  \bibinfo{author}{\bibfnamefont{T.}~\bibnamefont{Papenbrock}},
  \bibinfo{journal}{Phys. Rev. Lett.} \textbf{\bibinfo{volume}{84}},
  \bibinfo{pages}{4553} (\bibinfo{year}{2000}).

\bibitem[{\citenamefont{Percival}(1973)}]{Percival1973}
\bibinfo{author}{\bibfnamefont{I.}~\bibnamefont{Percival}},
  \bibinfo{journal}{J. Phys. B} \textbf{\bibinfo{volume}{6}},
  \bibinfo{pages}{L229} (\bibinfo{year}{1973}).

\bibitem[{\citenamefont{Berry}(1977)}]{Berry1977}
\bibinfo{author}{\bibfnamefont{M.~V.} \bibnamefont{Berry}},
  \bibinfo{journal}{J. Phys. A} \textbf{\bibinfo{volume}{10}},
  \bibinfo{pages}{2083} (\bibinfo{year}{1977}).

\bibitem[{not({\natexlab{b}})}]{noteTemp}
\bibinfo{note}{More generally, Ref.~\cite{ZelevinskyRep1996} has shown that
  different definitions of temperatures, associated with the microcanonical
  ensemble, single-particle occupation numbers, and information entropy,
  coincide in the chaotic regime.}

\end{thebibliography}
\end{document}